\documentclass{mem}
\usepackage{natbib}\usepackage{txfonts}\usepackage{balance}
\usepackage{graphicx}
\usepackage[a4paper,breaklinks,dvipdfm]{hyperref}
\idline{to appear}{1}
\begin{document}
\def\teff{$T\rm_{eff }$}
\def\kms{$\mathrm {km s}^{-1}$}
\def\gsim{\stackrel{>}{\scriptstyle \sim}}
\def\lsim{\stackrel{<}{\scriptstyle \sim}}\newcommand{\gppr}{\stackrel{>}{\scriptstyle \sim}}

\title{
Cen A as $\gamma$- and UHE cosmic-ray Source
}

   \subtitle{}

\author{
Frank M. \,Rieger
}

    \offprints{F.M. Rieger}
\institute{
Max-Planck-Institut f\"ur Kernphysik, Saupfercheckweg 1, 69117 Heidelberg, Germany,
 \email{frank.rieger@mpi-hd.mpg.de}
 }

\authorrunning{F.M. Rieger }

\titlerunning{Cen A as $\gamma$- and UHE cosmic-ray Source}

\abstract{
Cen A has been recently detected in the high-energy (HE) and very high energy (VHE) 
$\gamma$-ray domain by Fermi-LAT and H.E.S.S. We review the observed characteristics and 
suggest a scenario where the VHE emission originates from the vicinity of the black hole. Motivated 
by the possible association of some ultra-high energy (UHE) cosmic ray (CR) events with Cen A, 
we further analyze the acceleration efficiency for a number of a mechanisms (including shock and 
stochastic acceleration), showing that most of them (apart perhaps from shear) have serious difficulties 
in accelerating protons beyond a few $10^{19}$ eV.
}
\maketitle{}

\section{Introduction}

Cen~A is the nearest active galaxy, less than 4 Mpc away. It has a peculiar radio morphology with two 
jets emerging from its nucleus, and giant radio lobes stretching out to 250 kpc and extending over a 
8 x 4 degree field on the sky (for review, see e.g. \citealt{israel98}). VLBI studies have shown that Cen~A 
is a non-blazar source with its jet inclined at viewing angles \mbox{$i \gsim 50^{\circ}$} and characterized 
by moderate bulk flow speeds of \mbox{$u_j \simeq (0.1-0.5)$ c} only (\citealt{tin98,har03,mue11}). 
Cen~A harbors a supermassive black hole (BH) of mass $m_{\rm BH} = (0.5-3) \times 10^8 M_{\odot}$ 
(\citealt{mar06,neu07}). Given its estimated bolometric luminosity \mbox{$\sim10^{43}$} erg/s (\citealt{why04}), 
Cen~A is rather under-luminous and believed to accrete at sub-Eddington rates. If its inner disk would 
remain cooling-dominated (standard disk), accretion rates $\dot{m}\sim 10^{-3} \dot{m}_{\rm Edd}$ and 
equipartition magnetic field strengths close to the BH of $B_0=(2 L_b/r_g^2 c)^{1/2} \simeq 2 \times 
10^3$ G are expected ($r_g = GM/c^2$ is the gravitational radius). If the disk would switch to a radiatively 
inefficient (RIAF) mode, characteristic field strengths may reach $B_0 \sim 10^4$ G.

\section{Gamma-rays from Cen A}
Cen~A was the only non-blazar AGN detected at MeV to GeV energies by all instruments on board the 
Compton Gamma-Ray Observatory (for review, see \citealt{steinle10}). Fermi-LAT has recently reported 
HE ($>0.2$ GeV) $\gamma$-rays from both the giant radio lobes and the "core" (i.e., within \mbox{$\sim 
0.1^{\circ}$}) of Cen~A (\citealt{abdo10a,abdo10b}): Both lobes have been detected up to 3 GeV, with step 
spectral slopes (photon indices close to $2.6$), and contribute more than one-half to the total HE source 
emission. The HE lobe emission can be modeled as due to Compton up-scattering of CMB ($\epsilon=8 
\times 10^{-4}$ eV) and infrared extragalactic background photons by electrons with Lorentz factors $\gamma_e
=6\times 10^5$, assuming fields strengths $B \simeq 0.9\mu$G. This would imply a total energy (assuming a 
negligible proton contribution) in both lobes of \mbox{$E_t \simeq 10^{58}$} erg, and require a jet kinetic power 
\mbox{$L_j  \simeq 8 \times 10^{42}$ erg/s} close to the available accretion power. Fermi-LAT has also reported 
HE emission up to 10 GeV from the core, again with a steep photon index of \mbox{$\sim2.7$} and with apparent 
(isotropic) luminosity $L(>0.1\,\mathrm{GeV}) \simeq 4 \times 10^{40}$ erg/s. The HE light curve (using 15 d bins) 
is consistent with no variability.\\ 
VHE ($>0.1$ TeV) emission up to 5 TeV has been detected by H.E.S.S. in more than 100h of data taken in 
between 2004-2008 (\citealt{aha09}). The VHE spectrum is consistent with a power-law of photon index $2.7 \pm 
0.5$, and the apparent (isotropic) luminosity is $L(>250 \,\mathrm{GeV}) \simeq 2 \times 10^{39}$ erg/s. No 
significant variability has been found.\\
The nuclear SED of Cen~A, based on non-simultaneous data, shows two peaks, one at several 
$10^{13}$ Hz and one at around $0.1$ MeV (\citealt{chiaberge01,meisenheimer07,abdo10b}).
The SED below a few GeV is satisfactorily described by a one-zone synchrotron self-Compton (SSC) 
model (e.g., \citealt{chiaberge01}). As it turns out, however, the same approach fails to account for the 
TeV emission observed by H.E.S.S (\citealt{abdo10b}). In fact, a simple extrapolation of the Fermi (power 
law) spectrum tends to under-predict the observed TeV flux. This could indicate an additional contribution 
to the VHE domain beyond the conventional SSC jet emission, emerging at the highest energies. 
Non-thermal processes in the black-hole magnetosphere could offer a plausible explanation for this
(\citealt{rie09b}): Provided the inner disk in Cen~A is radiatively inefficient (ADAF-type), electrons can be
centrifugally accelerated along rotation magnetic field lines to $ \gamma_e \propto 1/(1-r/r_{\rm L}) \sim 
5 \times 10^7$ while approaching the light cylinder $r_{\rm L}=c/\Omega$, and thereby enable Compton 
up-scattering of sub-mm ADAF disk photons to the TeV domain, satisfying the observed VHE spectral 
and luminosity constraints. If the inner disk is of the ADAF-type, these VHE photons can also escape 
$\gamma\gamma$-absorption. Observationally, the nuclear SED of Cen~A peaks in the mid-infrared, 
with an apparent (isotropic) spectral luminosity of $\sim 6 \times 10^{41}$ erg/s at \mbox{$h\nu \sim 0.15$ 
eV} and evidence for an exponential cut-off (!) towards higher frequencies (\citealt{why04,meisenheimer07}). 
This emission is usually believed to be produced on larger scales, either by a non-thermal (non-isotropic!) 
synchrotron jet component at a distance $\gsim 0.03$ pc (e.g., \citealt{meisenheimer07}) or a (quasi-isotropic) 
dusty torus on scales $\sim 0.1$ pc or larger (e.g., \citealt{rad08}), thereby enabling sufficient dilution such 
that VHE photons are able to escape.

\section{UHE cosmic rays from Cen A}
The apparent clustering of UHECRs along Cen~A has renewed the interest into nearby AGN as potential 
UHECR accelerators. In 2007, the Pierre Auger (PAO) Collaboration initially reported evidence for an 
anisotropy at the 99\% confidence level in the arrival directions of cosmic-rays with energies $\gsim 6\times 
10^{19}$ eV (\citealt{PAO2007}). The anisotropy was measured by the fraction of arrival directions that were 
less than $\sim3^{\circ}$ from the positions of nearby AGN (within 75 Mpc) from the VCV catalog. While this 
correlation has become weaker given the now available (twice as large) data set, the updated analysis still 
suggests that a region of the sky around the position of Cen~A has the largest excess of arrival directions 
relative to isotropic expectations (\citealt{PAO2010a}). This obviously motivates a theoretical investigation of 
possible UHECR acceleration sites in Cen~A. Below we analyze the efficiency constraints expected for a 
number of acceleration mechanisms when applied to Cen~A. As it turns out, most mechanisms have serious 
difficulties in accelerating protons beyond a few $10^{19}$ eV. While the experimental situation is not fully 
conclusive yet, this result may fit into recent PAO indications for an increase of the average mass 
composition with rising energies up to $E\simeq 10^{19.6}$ eV (\citealt{PAO2010b}).

\subsection{CR acceleration in the BH vicinity}
\noindent
Rotating magnetic fields, either driven by the disk or the BH itself, could facilitate 
acceleration of charged particles:\\
{\bf (i) Direct electric field acceleration:} 
If the BH is embedded in a poloidal field of strength $B_p$ and rotating with angular frequency 
$\Omega_H$, it induces an electric field of magnitude $|\vec{E}| \sim (\Omega_H r_H) B_p/c$. 
This corresponds to a voltage drop across the horizon $r_H$ of magnitude $\Phi \sim r_H |\vec{E}|$.  
For Cen~A, this voltage drop becomes
\begin{equation}
  \Phi \sim 3 \times 10^{19} a \left( \frac{m_{\rm BH}}{10^8 M_{\odot}} \right) 
           \left(\frac{B_p}{10^4\mathrm{G}}\right)~~\mathrm{[V]}\,,
\end{equation} where \mbox{$0\leq a\leq 1$} denotes the dimensionless spin parameter. If a charged 
particle (charge number $Z$) could fully tap this potential, acceleration to $E=Z\,e\,\Phi \sim 3\times 
10^{19} Z$~eV may become possible. This would favor a rather heavy composition (e.g., irons instead 
of protons) above \mbox{$E_c = 5 \times 10^{19}$ eV}. 
Yet, whether such energies can indeed be achieved seems questionable: First, the charge density 
produced by annihilation of MeV photons emitted by an ADAF in Cen~A most likely exceeds the 
Goldreich-Julian (GJ) density required to screen the electric field (\citealt{levinson2011}). 
A non-negligible part of the electric field would then be no longer available for particle acceleration. 
Secondly, even if screening could be avoided, curvature losses (\citealt{lev00}) would constrain 
achievable energies for protons to 
\begin{equation}
  E_p \lsim 10^{19} a^{1/4} \left(\frac{M}{10^8M_{\odot}}\right)^{1/2}
       \left(\frac{B_p}{10^4\mathrm{G}}\right)^{3/4}~\mathrm{eV}.
\end{equation}
Thirdly, large-scale poloidal fields with strengths \mbox{$B_p \sim 10^4$ }G would be required. 
This seems overly optimistic, at least for a standard disk (\citealt{livio99}). Fourthly, one would need 
\mbox{$a\simeq1$} although rather moderate spins are expected for FR~I sources (\citealt{dal11}). 
Therefore, efficient DC acceleration of protons to $E_c$ and beyond in Cen~A is unlikely, but 
could be possible for heavier elements.\\
{\bf (ii) Centrifugal particle acceleration:}
Even if the charge density would exceed the GJ density, centrifugal particle acceleration along 
rotating magnetic field lines could still occur (e.g., \citealt{osm07,rie09b}). Yet, requiring that the 
acceleration timescale remains larger than the inverse of the relativistic gyro-frequency, CR 
Lorentz factors are limited to
\begin{equation}
  \gamma \lsim 2 \times 10^7 \gamma_0^{1/3} Z^{2/3} 
                  \left(\frac{m_p}{m_0}\right)^{2/3}
                  \left(\frac{r_{\rm L}}{10^{14}\mathrm{cm}}\right)^{2/3}\,
\end{equation} where $r_{\rm L}$ is the light cylinder radius (typically of a few $r_g$). 
This suggests that centrifugal acceleration is unable to produce UHECRs.
\subsection{Fermi-type CR acceleration in the jets and beyond}
Suppose instead that CR acceleration is Fermi-type, i.e., due to multiple scattering off moving 
magnetic inhomogeneities, with a small energy change in each scattering event. We may then 
distinguish the following scenarios:\\
{\bf (i) Diffusive shock (1st order Fermi):} 
In this case, energetic charged particles are assumed to pass unaffected through a shock front 
and, by being elastically scattered in the fluid on either side, to cross and re-cross it several times. 
Sampling the difference $\Delta u$ in flow velocities across a shock (always head-on), 
the characteristic energy gain for a particle crossing the shock, becomes 1st order, i.e., $\Delta 
\epsilon/\epsilon_1 \propto (\Delta u/c)$. As this is acquired during a shock crossing time $t_c \sim 
\lambda/u_s$ (with $u_s$ the shock speed and $\lambda$ the scattering mean free path), the 
characteristic acceleration timescale (for a non-relativistic shock) becomes
\begin{equation}\label{tshock}
t_{\rm acc} \simeq \frac{\epsilon}{(d\epsilon/dt)} \simeq \left(\frac{\epsilon_1}{\Delta\epsilon}\right) t_c 
                    \simeq \lambda \frac{c}{u_s^2}\,.
\end{equation} We can equate $t_{\rm acc}$ with the timescale for cross-field diffusion out of the 
system, $t_e\sim r_w^2/(\lambda c)$, or the dynamical timescale, $t_d \sim z/u_s$ (whichever is 
smaller), to obtain an estimate for the maximum achievable particle energy, $ E_{\rm max} \simeq 
Z e B r _w \beta_s$ (cf. \citealt{hil84}), i.e.
\begin{equation} 
 E_{\rm max} \simeq 2\times 10^{19} Z \left(\frac{B_0}{10^4\mathrm{G}}\right)
                        \left(\frac{\beta_s}{0.1}\right)~~\mathrm{eV}\,,
\end{equation} assuming \mbox{$\lambda \sim r_{\rm gyro}$}, with $r_{\rm gyro}$ the gyro-radius, 
$\beta_s =u_s/c$, and $B(z) \simeq 4~B_0~(r_g/z \alpha_j)$ for the typical magnetic field strength 
at location $z$ (allowing for magnetic field compression by a factor of $4$). Here, $\alpha_j$ is the jet 
opening angle. Radio observations of Cen~A indicate bulk flow speeds, both (!) on sub-pc and 
hundreds of pc scales, that are only mildly relativistic. This suggests only moderate internal shock 
speeds, $\beta_s \lsim 0.2$. Modest shock speeds are also supported by the nuclear SED of Cen~A 
with a synchrotron peak below $10^{20}$ Hz (cf. Lenain et al. 2008) as synchrotron-limited electron 
shock acceleration results in a (magnetic field-independent) peak at $\sim 3 \times 10^{19} (\beta_s/
0.1)^2$ Hz. Thus, efficient shock acceleration of protons to energies $E_c$ and beyond seems 
unlikely. Moreover, it can be shown that otherwise also a jet power well in excess of the one expected 
for Cen~A as an FR I-type source would be required (\citealt{rie09a}).\\
\noindent {\bf (ii) Stochastic 2nd order Fermi:} 
In general, the average energy gain due to scattering off randomly moving magnetic inhomogeneities 
(waves) is only second order, i.e., $\Delta\epsilon/ \epsilon_1 \propto (u/c)^2$. As the energy gain is 
acquired over a mean scattering time $t_s \sim \lambda/c$, the characteristic acceleration timescale 
is 
\begin{equation}\label{t2ndFermi}
 t_{\rm acc} \sim \frac{\epsilon}{(d\epsilon/dt)} \sim 
                               \left(\frac{c}{v_A}\right)^2 \frac{\lambda}{c}\,,
\end{equation} 
assuming that scattering is due to Alfv\'{e}n waves moving with \mbox{$u=v_A=B/\sqrt{4\pi\rho}$}. 
Neglecting radiative losses, particle energies are limited by escape via cross-field diffusion to
\[
  E  \lsim 2 \times 10^{19} Z \left(\frac{R}{100~\mathrm{kpc}}\right)
                          \left(\frac{v_A}{0.1~c}\right)
                           \left(\frac{B}{1 \mu\mathrm{G}}\right) \mathrm{eV}
\] on scales of $R\sim 100$ kpc, appropriate for the giant radio lobes in Cen~A. 
Stochastic UHE proton acceleration in its lobes (as suggested in Hardcastle et al. 2009) would thus 
require Alfv\'{e}n speeds $v_A \geq 0.3$ c. This seems difficult to achieve (cf. also \citealt{osu09}). 
Thermal X-ray emission from the lobes \citealt{iso01} suggest (thermal) plasma densities of 
\mbox{$n_{\rm th} \simeq (10^{-5} - 10^{-4})$ cm$^{-3}$}, implying $v_A \lsim 0.003$ c. 
Such values for $n_{\rm th}$ are consistent with independent estimates based on Faraday rotation 
measurements (\citealt{fea09}). Efficient UHECR acceleration in the lobes seems thus rather unlikely.\\
\noindent {\bf (iii) Shear acceleration:}
If the flow profile is non-uniform across the jet, as in the case of a shear flow with $\vec{u}=u_z(r)
\vec{e}_z$, then energetic particles scattered across it, may be able to sample the flow difference 
$du$ and thereby get accelerated (\citealt{jok90,rie06}). Like stochastic 2nd order Fermi, the average 
energy gain is $\propto (du/c)^2$, although the physical origin is different (i.e., due to the systematic, 
instead of the random motion of the scatterers). The velocity difference, that a particles experiences, 
is $du \sim (du_z/dr)\lambda$, where $\lambda$ is the scattering mean free path. Again, this energy 
change is acquired over $\tau_s \sim \lambda/c$, so that
\begin{equation}
  t_{\rm acc} \sim \frac{\epsilon_1}{\Delta\epsilon/\tau_s} 
                      \sim \frac{1}{(du_z(r)/dr)^2}\frac{c}{\lambda}\,.
\end{equation} In contrast to eq.~(\ref{tshock}) and eq.~(\ref{t2ndFermi}), now $t_{\rm acc} \propto
1/\lambda$. Thus, $t_{\rm acc}$ becomes smaller as a particle increases its energy (so that its 
$\lambda \propto p^{\alpha}$, $\alpha>0$, becomes larger). Shear acceleration therefore preferentially 
picks up high-energy seed particles for further energization, and acts more easily on particles of higher 
rigidity. Shocks, operating in the jet, could well provide the required seed particles (\citealt{rie09b}). 
Achievable particle energies are then constrained by the confinement condition (i.e., $r_{\rm gyro} \leq$ 
width of the shear layer). The large-scale jet in Cen~A has a projected length of about $4.5$ kpc, and 
towards its end a width of about $1$ kpc (\citealt{bur83,kra02}). For a characteristic $B\sim 10^{-4} b_j$ G 
on kpc-scale (cf. magnetic flux conservation) and a width of the shear comparable to the width of the jet, 
maximum energies 
\begin{equation}
E \sim Z e B (\Delta r) \sim 10^{20} b_j Z ~\mathrm{eV}\,
\end{equation} are possible. Shear acceleration might thus be able to boost energetic seed protons 
(e.g., produced by shock acceleration) to energies beyond $E_c$. If the magnetic field gets amplified 
by internal shear (e.g., \citealt{urp06}), even $b_j \gsim 1$ may be possible. Note that a shear dynamo 
could possibly explain why the magnetic field direction in Cen~A seems to be almost parallel along 
its kpc jet (\citealt{har03}).

\section{Conclusions}
The observed HE (Fermi) and VHE (H.E.S.S.) characteristics of Cen~A ("core") suggest 
that the HE and VHE emission originate from different regions. While the nuclear SED below 
a few GeV can be satisfactorily described with a conventional one-zone SSC model, an 
additional contribution is required to account for the VHE emission. Non-thermal processes 
in the black hole-jet magnetosphere could offer a plausible explanation for the latter.\\  
Whether Cen~A is indeed an UHECR source has observationally not been settled yet. From 
a theoretical point of view, efficient acceleration of protons to UHECR energies in Cen~A 
remains challenging in the framework of most standard mechanisms. Observational evidence 
for UHE protons may therefore support the operation of an additional acceleration mechanism 
("two-step") such as shear. The situation is much more relaxed for heavier elements like iron 
nuclei, which could most likely be directly accelerated (either by shocks or within the BH 
magnetosphere) to UHECR energies. Note that simultaneous operation of several mechanisms 
(with maximum energy not always linearly dependent on charge) also seems to constrain the 
potential to infer the UHECR composition from the observed anisotropy (\citealt{lem09}).

\begin{acknowledgements}
I am grateful to Franco Giovannelli for the invitation to give this talk.
\end{acknowledgements}

\bibliographystyle{aa}

\bigskip
\noindent {\bf DISCUSSION:}
\bigskip

\noindent {\bf VALENTI BOSCH-RAMON:} Assuming ideal MHD in the jet launching region, how high can 
the particle Lorentz factor be?\\ 

\noindent {\bf FRANK RIEGER:} In ideal (single-fluid) MHD models, the azimuthal component of the magnetic field 
increases on approaching the light cylinder, and this makes centrifugal (test) particle acceleration 
inefficient. The details are dependent on the assumed rotation law and current distribution along the 
field line. It is, however, not always clear whether ideal MHD can be applied close to the BH (cf. Levinson
 \& Rieger 2011).

\end{document}